\documentclass[nature,article,twocolumn,superscriptaddress,floatfix]{revtex4-2}
\usepackage[utf8]{inputenc}
\usepackage{bm}
\usepackage[pdftex]{graphicx}
\usepackage{multirow,amssymb,amsbsy,amsmath}
\usepackage{threeparttable}
\usepackage{verbatim}
\usepackage{color}
\makeatletter
\if@twocolumn
\fi
\makeatother
\usepackage{hyperref}
\usepackage[T1]{fontenc}
\usepackage{natbib}
\usepackage{booktabs}
\usepackage{diagbox}
\usepackage{pifont}
\usepackage[labelsep=none]{caption}
\makeatletter
\renewcommand{\section}{%
	\@startsection{section}{1}{\z@}%
	{-3.5ex \@plus -1ex \@minus -.2ex}%
	{2.3ex \@plus.2ex}%
	{\normalfont\Large\bfseries}}
\renewcommand{\subsection}{%
	\@startsection{subsection}{2}{\z@}%
	{-3.25ex \@plus -1ex \@minus -.2ex}%
	{1.5ex \@plus .2ex}%
	{\normalfont\large\bfseries}}
\makeatother

\begin{document}
\captionsetup[figure]{justification=raggedright,labelfont={bf},name={Fig. }}
\title{Storage of telecom-band time-bin qubits in thin-film lithium niobate}
\author{Xiao-Jie Wang}
\altaffiliation{These authors contributed equally to this work.}
\author{Yong-Teng Wang}
\altaffiliation{These authors contributed equally to this work.}
\author{Zi-Wei Zhao}
\affiliation{State Key Laboratory of Quantum Optics Technologies and Devices, Institute of Opto-Electronics, Shanxi University, Taiyuan, 030006, China}
\author{Yong-Min Li}
\email{yongmin@sxu.edu.cn}
\affiliation{State Key Laboratory of Quantum Optics Technologies and Devices, Institute of Opto-Electronics, Shanxi University, Taiyuan, 030006, China}
\affiliation{Collaborative Innovation Center of Extreme Optics, Shanxi University, Taiyuan, 030006, China}
\author{Tian-Shu Yang}
\email{tianshuyang@sxu.edu.cn}
\affiliation{State Key Laboratory of Quantum Optics Technologies and Devices, Institute of Opto-Electronics, Shanxi University, Taiyuan, 030006, China}
\date{\today}
\maketitle
{\textbf{Integrated photonics has emerged as a promising platform for quantum communication and quantum computation. Thin-film lithium niobate (TFLN) has gained significant attention in this field due to its exceptional optical properties, enabling the realization of numerous integrated photonic devices. However, quantum memory, which serves as a universal building block for the quantum internet, has not yet been demonstrated in TFLN. In this study, we realized the first on-chip quantum memory using erbium ions doped TFLN. The developed quantum memory achieves a storage time of 400 ns with an efficiency of 1.95\% \(\pm\) 0.04\%, significantly outperforming conventional waveguide delay lines. The multimode capability is demonstrated by successfully storing four temporal modes. Furthermore, single-photon-level coherent pulses are encoded into time-bin qubits and stored with a fidelity of 96.8\% \(\pm\) 0.3\%, surpassing the classical limit achievable by measure-and-prepare strategy. Our results demonstrate the first on-chip quantum memory for telecom-band time-bin qubits in TFLN, providing a key building block toward integrated quantum registers and repeaters for scalable quantum information processing.}}

\section*{Introduction}
Integrated quantum photonics has established itself as a key platform for advancing quantum information technologies \cite{Wang2020,Pelucchi2022,Moody_2022,PRXQuantum2024}. By enabling the on-chip integration of diverse optical components, it offers superior scalability, lower power consumption, smaller device footprint, and improved cost-effectiveness compared with bulk optics. 
A variety of material platforms have been investigated for integrated photonics, including silicon-based materials (such as Si, SiC, SiN), III-V and III-N semiconductors (e.g., GaAs, InP, AlN, GaN), and ferroelectrics (LiNbO$_{3}$ and LiTaO$_{3}$) \cite{Moody_2022,PRXQuantum2024,Elshaari2020,lnreview_1,LiTaO2024,LiTaO2025}. 
Among these material platforms, thin-film lithium niobate (TFLN) is particularly promising for integrated photonics because of its remarkable properties, including large nonlinear coefficients, strong electro-optic effect, and broad transparency window \cite{lnreview_1,ScienceRev2023,TFLN2025}.
These attributes have enabled the realization of key integrated quantum photonic devices on TFLN, such as low-voltage electro-optic modulators \cite{cmos-eom}, ultra-low threshold frequency combs \cite{ofc2019,OFC2025}, ultrabroadband entangled photon-pair sources \cite{Entangled2021}, and microwave photonic signal processing engines \cite{WangCheng2024}. 
However, the realization of a photonic quantum memory on TFLN remains an outstanding challenge \cite{Luo2023}.

Quantum memory, which enables the storage and retrieval of quantum states, is critical for advancing quantum information technologies \cite{Pelucchi2022,PRXQuantum2024,Elshaari2020}. 
It serves as a foundational element not only for quantum repeaters in long-distance quantum networks to mitigate fiber loss \cite{science2018,RMP2023}, but also for quantum registers within quantum computing architectures to reduce qubits requirements \cite{QC2021}. 

Erbium ions (Er$^{3+}$) doped in TFLN represent a promising candidate for realizing on-chip quantum memory \cite{ErTFLN2022,ZhouQiang2023,PRApplied2019,PRL2023ERLN}. 
Erbium ions exhibit long optical coherence times in lithium niobate (LN), with absorption and emission transitions wavelength at telecom band (around 1532 nm) \cite{ErLN2010}. Quantum memories based on bulk Er$^{3+}$:LN waveguides have already been demonstrated \cite{ZhouQiang2023,PRApplied2019,AFCZeman2022}, while recent advances have also achieved atomic frequency comb (AFC) memories based on Tm$^{3+}$ doped TFLN \cite{nanolett2020,AFC2023}. Furthermore, the spectroscopic properties of Er$^{3+}$ ions in TFLN have recently been investigated \cite{ErTFLN2022}.  
To date, a quantum memory directly implemented in Er$^{3+}$:TFLN that is compatible with telecommunication-wavelength photons has not yet been achieved.

In this work, we realized an on-chip photonic quantum memory based on Er$^{3+}$:TFLN. Employing the AFC protocol, we achieve a storage time of 400 ns with an efficiency of 1.95\% \(\pm\) 0.04\%. By leveraging the intrinsic temporal multimode capability of the AFC scheme, storage of four temporal modes is realized. Moreover, the input signals were encoded as time-bin qubits, achieving a memory fidelity up to 96.8\% \(\pm\) 0.3\% for an input mean photon number of \(\mu_{in}\) = 0.578, which clearly exceeds the classical benchmark associated with measure-and-prepare strategies. These results pave the way for the implementation of TFLN in integrated quantum technologies.

\section*{Results}
\subsection*{Device preparation} 
\begin{figure*}
\centering
\includegraphics[width=1\textwidth]{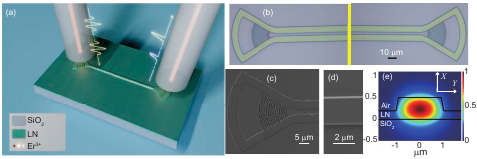}
\caption{\label{Fig:1} (a) Schematic of the integrated quantum memory on the TFLN platform. Quantum states are coupled into and out of the waveguide through grating couplers, the waveguide is doped with Er$^{3+}$ ions. (b) Optical microscope image of the fabricated grating couplers and the 5 mm-long straight waveguide. The yellow-shaded region is not shown to scale. (c), (d) Scanning electron microscope image of the grating coupler and waveguide. (e) \textcolor{red}{Finite-difference time-domain (FDTD) simulation of the fundamental transverse-electric (TE) mode profile at 1531.6 nm in the waveguide, showing an overlap of 85.37\% between the optical mode and the Er$^{3+}$-doped TFLN region.} The \textit{X}–\textit{Y} plane represents the LN crystal axes, while the waveguide is oriented along the \textit{Z}-axis.}
\end{figure*}
The schematic of the photonic quantum memory implemented on the TFLN chip is shown in Fig. \ref{Fig:1}(a). The quantum memory is implemented in an Er$^{3+}$ doped waveguide, which can be integrated with other components on the same chip or coupled to optical fibers through the grating couplers. In this study, the Er$^{3+}$:TFLN was fabricated on a commercially available \textit{x}-cut TFLN wafer (Nanoln) with a thickness of 500 nm, doped with 0.5 mol\% Er$^{3+}$ ions. The Er$^{3+}$:TFLN layer was bonded onto a 4.7 $\mu$m silica layer, which was grown on a 525 $\mu$m thick silicon substrate. The waveguide has a length of 5 mm, with light coupled in and out through two grating couplers, as shown in Fig. \ref{Fig:1}(b). The scanning electron microscope images of the grating coupler and waveguide are shown in Fig. \ref{Fig:1}(c) and \ref{Fig:1}(d). The coupling efficiency of the grating coupler is about 12.5\% at 1531.6 nm, and the corresponding coupling efficiency curve is shown in Supplemental Fig. S2. Figure \ref{Fig:1} (e) shows a finite-difference time-domain simulation of the electric field profiles in the waveguides for the fundamental transverse-electric mode at 1531.6 nm.

\subsection*{Spectroscopic properties of Er$^{3+}$:TFLN} 
\begin{figure*}
\centering
\includegraphics[width=1\textwidth]{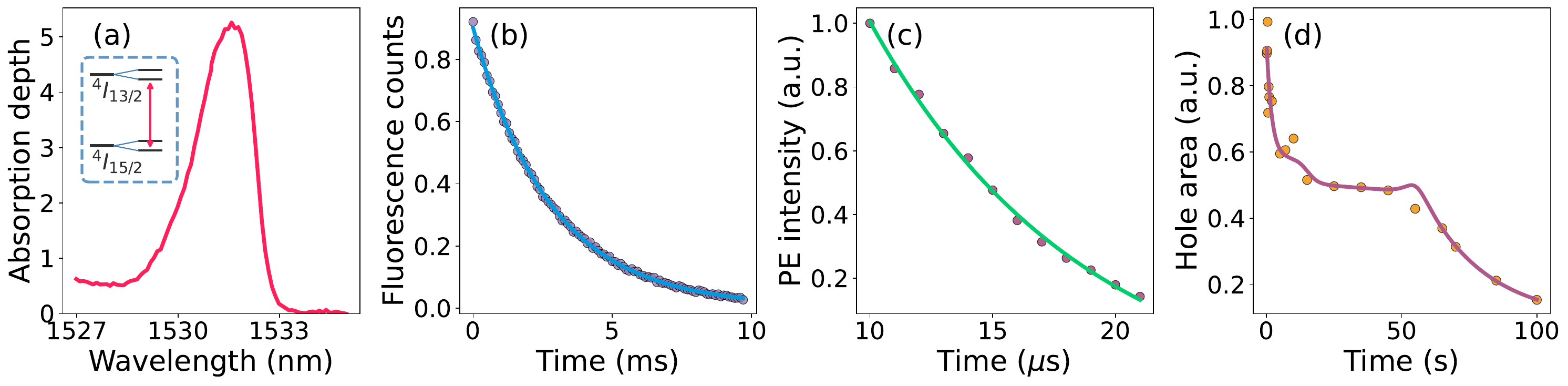}
\caption{\label{Fig:2} (a) Absorption spectrum of Er$^{3+}$ ions in a 5 mm long TFLN waveguide. The inset shows the simplified energy-level diagram of Er$^{3+}$ ions in TFLN. (b) Fluorescence decay of the $^{4}$\textit{I}$_{13/2}$ state at 1.6 K, with a single exponential fit (blue line) yielding a lifetime of T$_{1}$ = 2.78 $\pm$ 0.02 ms. (c) Two-pulse photon echo (PE) decay of the $^{4}$\textit{I}$_{15/2}$ \(\leftrightarrow \) $^{4}$\textit{I}$_{13/2}$ transition, fitted with an exponential function (green line) giving a coherence time of T$_{2}$ = 17.48 \(\pm\) 1.69 $\mu$s. (d) Decay of the spectral hole area, fitted using three exponential decay functions. The shortest hole lifetime is estimated to 1.95 \(\pm\) 1.09 s.} 
\end{figure*}

Before realizing quantum memory based on Er$^{3+}$:TFLN, it is essential to characterize the spectroscopic properties of the Er$^{3+}$ ions. The relevant energy-level structure of Er$^{3+}$ in TFLN used in this study is shown in the inset of Fig. \ref{Fig:2}(a). To achieve a long optical coherence time, the device is operated at cryogenic temperatures to suppress phonon-induced decoherence, and a strong magnetic field is applied to slow the Er$^{3+}$ spin-flip process \cite{ErLN2010,ErYSO1s}. Consequently, the TFLN chip is placed in a cryogenic environment at 1.6 K and 4 T using an Oxford SpectromagPT system. The detailed experimental setup is presented in Supplemental Fig. S1. The light polarization is aligned along the \textit{Y} axis (\(E\parallel Y\)), while the magnetic field is applied along the \textit{Z} axis (\(B\parallel Z\)). The coordinate axes (\textit{X}, \textit{Y}, and \textit{Z}) are defined as shown in Fig. \ref{Fig:1}(e). 

First, we measured the absorption properties of the $^{4}$\textit{I}$_{15/2}$ \(\leftrightarrow \) $^{4}$\textit{I}$_{13/2}$ transition. The corresponding absorption spectrum is presented in Fig. \ref{Fig:2}(a). The peak absorption coefficient of 10.52 cm\(^{-1}\) is observed at 1531.6 nm. The results indicate an optical inhomogeneous linewidth of 1.97 nm (250 GHz). 
Then, we measured the fluorescence lifetime (optical T$_1$) of the $^{4}$\textit{I}$_{13/2}$ excited state. A single frequency laser pulse with a duration of 100 \(\mu\)s was used to excite the $^{4}$\textit{I}$_{15/2}$ \(\leftrightarrow \) $^{4}$\textit{I}$_{13/2}$ transition, and the corresponding fluorescence decay was recorded. The measured optical T$_1$ is 2.78 $\pm$ 0.02 ms, as shown in Fig. \ref{Fig:2}(b), which is consistent with previous reports \cite{ErLN2010, ErTFLN2022,Erions2023}. Since photonic quantum memories rely on optical coherence, we further measured the optical coherence time (T$_2$) of the $^{4}$\textit{I}$_{15/2}$ \(\leftrightarrow \) $^{4}$\textit{I}$_{13/2}$ transition using the photon echo. The measured results are shown in Fig. \ref{Fig:2}(c), from which a coherence time of 17.48 \(\pm\) 1.69 $\mu$s is obtained. This value is shorter than that reported for Er$^{3+}$ ions in bulk LN under comparable experimental conditions \cite{ErLN2010}. The reduction in T$_2$ is attributed to the higher concentration of dopants in our TFLN.

As Er$^{3+}$:TFLN represents a potential platform for implementing on-chip quantum memory, efficient ground-state population manipulation, such as spectral hole burning and the creation of persistent absorption features, is essential. 
The realization of these processes requires that the ground-state $^{4}$\textit{I}$_{15/2}$ lifetime significantly exceeds that of the excited state $^{4}$\textit{I}$_{13/2}$. 
To investigate the ground-state lifetime, we measured the temporal decay of the spectral hole area, as shown in Fig. \ref{Fig:2}(d). The shortest decay time obtained from the fit is 1.95 \(\pm\) 1.09 s. This spectral hole lifetime is approximately 700 times longer than the excited-state lifetime (2.78 $\pm$ 0.02 ms), confirming that the ground state is sufficiently long-lived to enable efficient optical pumping.
\begin{figure*}
\centering
\includegraphics[width=1\textwidth]{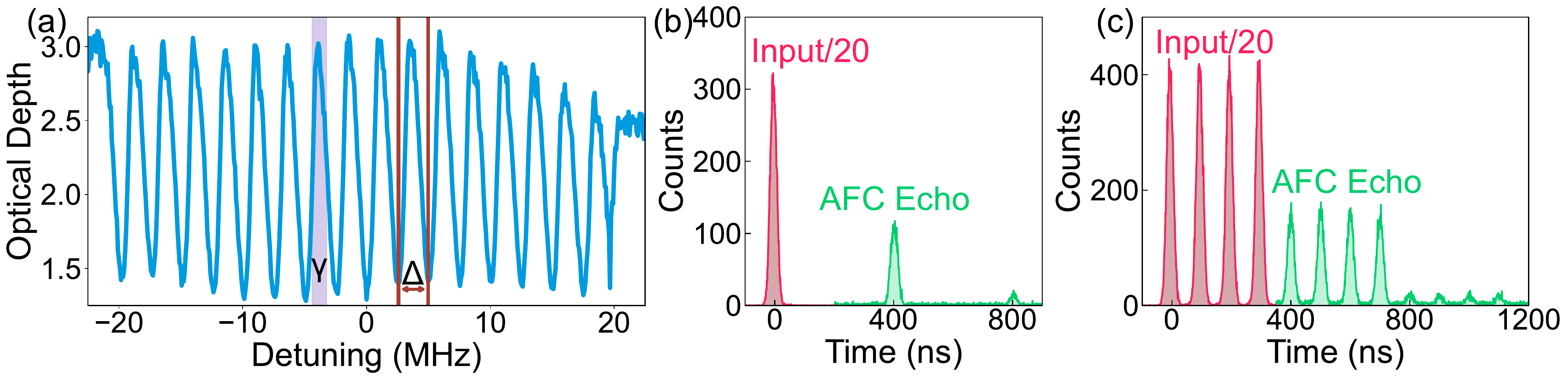}
\caption{\label{Fig:3} (a) AFC structure created in Er$^{3+}$:TFLN chip via spectral hole burning. The teeth spacing \(\Delta\) is 2.5 MHz, corresponding to storage time of 400 ns. The linewidth of the teeth is \(\gamma\) = 1.03 \(\pm\) 0.02 MHz. (b) Time histograms of the input photons (red) and the photons retrieved after a storage time of 400 ns (green), measured at an input mean photon number of \(\mu_{in}\) = 0.578. The bin size is 1 ns in this histogram. (c) Time histograms of four temporal modes memory.}
\end{figure*}
\subsection*{Atomic frequency comb storage} 

To store the telecom-band photons in Er$^{3+}$:TFLN chip, we employ the AFC scheme \cite{AFC2008,AFC2009}. The AFC scheme relies on spectral hole burning techniques to periodically tailor the inhomogeneously broadened absorption profile of Er$^{3+}$ ions into a series of narrow, equally spaced absorption peaks, as illustrated in Fig. \ref{Fig:3}(a). 
When the input photons are absorbed by the comb, they will be re-emitted after a fixed delay time \(\tau\) = 1/\(\Delta\), where \(\Delta\) denotes the comb spacing.
As shown in Fig. \ref{Fig:3}(a), we prepared a 40 MHz-wide AFC with a tooth spacing of 2.5 MHz, corresponding to a memory time of \(\tau\) = 400 ns. 
The time histogram of the memory for \(\mu_{in}\) = 0.578 is shown in Fig. \ref{Fig:3}(b). The observed memory efficiency is 1.95\% \(\pm\) 0.04\%, with a signal-to-noise ratio of 56.3 \(\pm\) 7.0. The dominant noise contribution arises from detector dark counts.

The AFC protocol inherently supports temporal multimodes, where the maximum number of storable temporal modes is determined by the ratio of the mode duration to the storage time \cite{1250modes}. To verify the multimode capacity of our storage, we input a sequence of four pulses. The output signals were retrieved in a first-in-first-out fashion, as evidenced by the measurements in Fig. \ref{Fig:3}(c). The average memory efficiency is 1.97\% \(\pm\) 0.02\%. This temporal multiplexing capability provides a scalable means to boost the entanglement distribution rate, thereby improving the overall efficiency of quantum-repeater-based networks \cite{RMP2023,RMP2011,wvv1-6lg8}.

\subsection*{Time bin qubits storage}
\begin{figure*}[tb]
\centering
\includegraphics[width=1\textwidth]{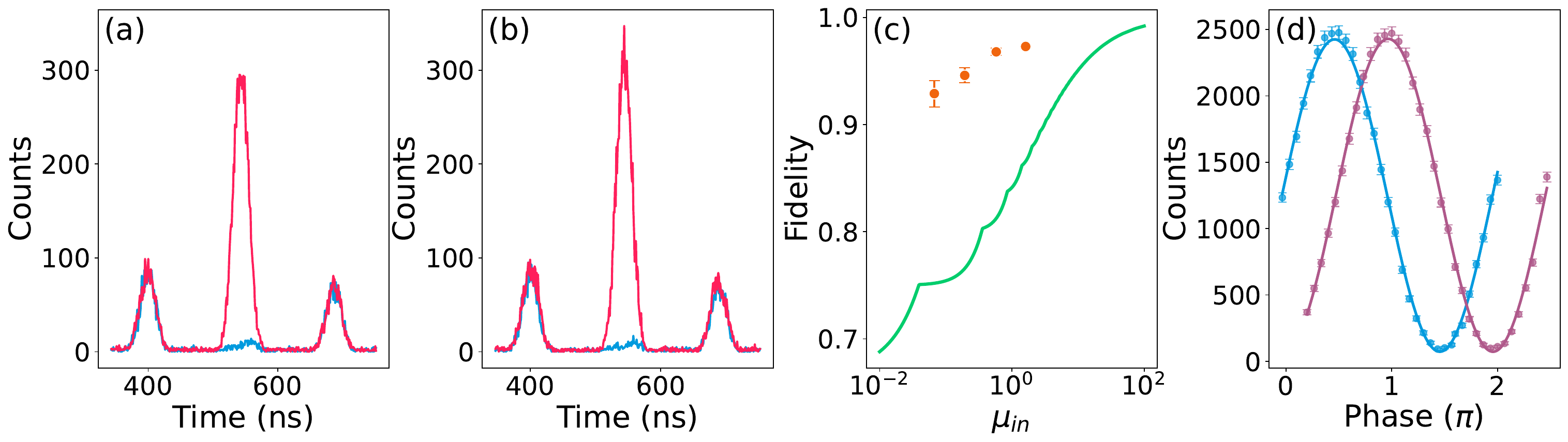}
\caption{\label{Fig:4} (a), (b) Photon-counting histograms for the retrieved states \(\lvert e \rangle + \lvert l \rangle\) and
\(\lvert e \rangle - \lvert l \rangle\), where constructive (destructive) interference is shown in red (blue). The interference peak corresponds to the central peak, and the average input photon number is \(\mu_{in}\) = 1.610. (c) Total fidelity as a function of the photon number per input qubit. The fidelity achieved by the classical measure-and-prepare strategy is indicated by the green line (see Supplementary Eq.~(2) for details). The orange points show the fidelities obtained with the memory; the full dataset is summarized in Supplementary Table S1. (d) Interference fringes of the central peak as a function of the relative phase \(\Delta\alpha\). Blue (purple) points denote the projections onto the \(\lvert e \rangle + \lvert l \rangle\) (\(\lvert e \rangle - i\lvert l \rangle\)) analysis states.}
\end{figure*}
Leveraging this temporal multimode capacity, we encode the input pulses as time-bin qubits to illustrate the potential of this memory for quantum technologies \cite{ZZQ2022,SWPRL2015}. This encoding scheme is widely employed in quantum networks owing to its robustness against depolarization and spatial mode mixing in long-distance optical fibers \cite{Timebin2025,timebin20252}. In our setup, the time-bin qubit \(\lvert e \rangle + e^{i\Delta\alpha} \lvert l \rangle\) is generated using an acousto-optic modulator (AOM) driven by an arbitrary waveform generator, where the relative phase \(\Delta\alpha\) is actively controlled. Here, \(\lvert e \rangle\) and \(\lvert l \rangle\) denote the \textit{early} and \textit{late} time-bin states, respectively. Each time-bin qubit consists of two temporal modes with a pulse width of 50 ns and a separation of 130 ns. Five time-bin qubits \(\lvert e \rangle\), \(\lvert l \rangle\), \(\lvert e \rangle \pm \lvert l \rangle\), and \(\lvert e \rangle + i \lvert l \rangle\) were used to characterize this memory device. At first, we evaluate the fidelity of \(\lvert e \rangle\) and \(\lvert l \rangle\), as shown in  Supplemental Fig. S5. For \(\mu_{in}\) = 1.610, the observed average fidelity value is \(F_{el}\) = 98.8\% \(\pm\) 0.1\%. To characterize the superposition of the time-bin qubits  \(\lvert + \rangle\) = \(\lvert e \rangle + \lvert l \rangle\) and  \(\lvert - \rangle\) = \(\lvert e \rangle -  \lvert l \rangle\), a custom-built unbalanced Mach–Zehnder interferometer (UMZI) was employed (as shown in Supplemental Fig. S3) \cite{timebin20252}. Figure \ref{Fig:4}(a) and \ref{Fig:4}(b) show the interference results for \(\mu_{in}\) = 1.610, from which a fidelity of \(F_{+-}\) = 96.6\% \(\pm\) 0.2\% is obtained. Therefore, the total conditional fidelity of the retrieved qubit \(F_{T} = \frac{1}{3} F_{el} + \frac{2}{3} F_{+-}\) is found to be 97.3\% \(\pm\) 0.2\%. This result is significantly higher than the maximal fidelity that can be achieved by the classical measure-and-prepare strategy \cite{Fidelity2011,SWPRL2015,Yang2018}, as illustrated in Fig. \ref{Fig:4}(c), demonstrating the genuine quantum nature of our storage device. 

To investigate the phase coherence of the stored time-bin qubits, we stabilize the UMZI to the \(\lvert e \rangle + \lvert l \rangle\) and \(\lvert e \rangle - i \lvert l \rangle\) states. By varying the phase \(\Delta\alpha\) from 0 to 2\(\pi\), a series of input qubit states are prepared. The qubits are stored in the memory, then retrieved and sent to the UMZI for analysis. The interference fringes are presented in Fig. \ref{Fig:4}(d). By applying sinusoidal fit of the form \cite{timebin20252}: \(Counts(\Delta\alpha) = Counts_{avg}(1+Vsin(\Delta\alpha))\), where \(Counts_{avg}\) denotes the average of the maximum and minimum counts,
we extract raw mean visibilities of 94.0\% \(\pm\) 0.4\% and 94.2\% \(\pm\) 0.4\%, which result in state fidelities of 97.0\% \(\pm\) 0.2\% and 97.1\% \(\pm\) 0.2\%, respectively. 
The obtained fidelities clearly exceed the classical limit for measure-and-prepare strategies, further confirming the quantum nature of the memory.

\section*{Discussion}
In conclusion, we have demonstrated a photonic time-bin qubits memory in a Er\(^{3+}\):TFLN chip. The implemented memory achieves a storage time of 400 ns with an efficiency of 1.95\% \(\pm\) 0.04\% (approximately –17.5 dB) in a 5 mm-long waveguide. To implement an equivalent time delay with an optical delay line, a 60-m waveguide is required, which would introduce a total attenuation of approximately -78 dB, even with an ultra-low loss of 1.3 dB/m \cite{1.3dB}. Compared with the waveguide delay line, the presented on-chip quantum memory provides distinct advantages for realizing long-time delays, offering substantially lower loss and significantly reduced footprint.

In the future, the memory efficiency can be improved by employing an impedance-matched cavity \cite{Cavity-Enhanced2025,CavityAFC2013}. While the memory lifetime can be further extended by operating at lower temperatures and stronger magnetic fields \cite{ErYSO1s}. It is worth noting that a more precise preparation of the AFC will lead to both higher storage efficiency and longer storage lifetime \cite{1250modes}. Furthermore, while the AFC scheme is a pre-programmed delay, on-demand readout can be realized by employing the Stark-modulated AFC protocol \cite{ZZQ2022,SAFC2021}. 

Looking forward, the integration of on-chip quantum memory units with low-loss waveguides \cite{1.3dB}, high-performance electro-optic modulators \cite{9-cry-lowvlotage}, high-quality entangled photon sources \cite{Entangled2021}, and single-photon detectors \cite{SNSPD2021} will further advance the functionality and scalability of the TFLN photonic platform. 
The compatibility of the memory with the telecom band makes this chip a promising candidate for long-distance quantum networking. Its integrated multi-functionalities may ultimately enable chip-scale quantum repeaters, thereby serving as a universal building block for telecom-band quantum communication and distributed quantum computing \cite{Liuxiao2024}.
\section*{Methods}
\subsection*{Fabrication of TFLN}
This device is fabricated in a 0.5 \%mol Er\(^{3+}\) doped \textit{x}-cut TFLN, where Er\(^{3+}\) ions are doped in the LN crystal growth process. The TFLN is prepared from bulk erbium doped lithium niobate by the SmartCut technique in Nanoln. Then, the TFLN is died to 10 mm\(\times\)12 mm (\textit{Y} \(\times\) \textit{Z}) pieces for further process. For the fabrication of the LN waveguides and grating couplers, a 200 nm-thick chromium layer was deposited as a hard mask. Subsequently, a 400 nm AR-P 6200 resist was spin-coated, and the device structures were defined using electron-beam lithography. The patterns were then transferred into the TFLN layer by argon-based dry etching with an etching depth of 300 nm. The resulting sidewall angle was approximately 70\(^\circ\). The SEM images of the fabricated structures are shown in Fig. \ref{Fig:1}(c) and \ref{Fig:1}(d). To facilitate convenient operation in cryogenic environments, the chip was packaged with two fiber arrays \cite{Yongten2025}.
\subsection*{AFC preparation and memory efficiency}
The absorption spectrum of Er$^{3+}$ ions is tailored to prepare the AFC using spectral hole burning, thereby forming well-defined absorption peaks. These peaks are created using a parallel preparation method \cite{PhysRevA2016}. Considering the optical lifetime T\(_{1}\), the preparation pulse has a length of 5 ms and is repeated 150 times. 
The created AFC structure is shown in Fig. \ref{Fig:3}(a). For Gaussian peaks of the AFC, the memory efficiency can be expressed as \cite{AFC2009}:
\begin{align}
\eta_{AFC,th} & = (\frac{d_{1}}{F} )^{2}e^{-d_{1}/F}e^{-7/F^{2}}e^{-d_{0}}
\end{align}
where $d_{1}$ = 1.61 \(\pm\) 0.03 is the comb optical depth, \textit{F} = \(\Delta/\gamma\) =  1.92 \(\pm\) 0.02 is finesse of the AFC, \(d_{0}\) = 1.36 \(\pm\) 0.01 is the absorbing background. The theoretical AFC efficiency is \(\eta_{AFC,th}\) = 1.41\% \(\pm\) 0.01\%. 
The prediction value of \(\eta_{AFC,th}\) is lower than the measured memory efficiency.

\bibliography{reference}

\section*{Data availability} The data that support the findings of this study are available from the corresponding authors on request. 

\section*{Acknowledgments}
This work was supported by the Fundamental Research Program of Shanxi Province (202403021211096) and supported by the National Natural Science Foundation of China (62105321).

\section*{Author contributions}
T.S.Y. designed experiment. T.S.Y., X.J.W and Y.T.W carried out the experiment assisted by Z.W.Z. T.S.Y., X.J.W., and Y.T.W. wrote the paper with input from other authors. T.S.Y. and Y.M.L supervised the project. All authors discussed the experimental procedures and results.

\section*{Additional information}
\textbf{Competing interests:} The authors declare no competing interests.

\end{document}